\documentclass{emulateapj}

\usepackage{amsmath,amssymb}

\usepackage{xcolor}

\slugcomment{}

\usepackage[bookmarksnumbered,
colorlinks,
linkcolor=blue,
citecolor=black,
filecolor=black,
urlcolor=blue,
breaklinks=true,
]{hyperref}
\newcommand{\kpc}{\,{\rm kpc}}

\newcommand{\magsec}{\,{\rm mag\,arcsec^{-2}}}
\newcommand{\Dmin}{{\,D_{\rm min}}}
\newcommand{\rmin}{{\,R_{\rm min}}}

\newcommand{\beqn}{\begin{equation}}
\newcommand{\eeqn}{\end{equation}}

\shortauthors{Liu et al.}
\begin{document}
\title{Serendipitous discovery of a massive $\lowercase{c}$D galaxy at $z=1.096$: Implications for the early formation and late evolution of $\lowercase{c}$D galaxies}
\author{F. S. Liu \altaffilmark{1,2},
Yicheng Guo \altaffilmark{2},
David C. Koo \altaffilmark{2},
Jonathan R. Trump \altaffilmark{2},
Guillermo Barro \altaffilmark{2},
Hassen Yesuf \altaffilmark{2},
S. M. Faber \altaffilmark{2},
M. Giavalisco \altaffilmark{3},
P. Cassata\altaffilmark{4},
A. M. Koekemoer\altaffilmark{5},
L. Pentericci \altaffilmark{6},
M. Castellano \altaffilmark{6},
Edmond Cheung \altaffilmark{2},
Shude Mao \altaffilmark{7},
X. Y. Xia \altaffilmark{8},
Norman A. Grogin \altaffilmark{5},
Nimish P. Hathi \altaffilmark{9},
Kuang-Han Huang \altaffilmark{10},
Dale Kocevski \altaffilmark{11},
Elizabeth J. McGrath \altaffilmark{12},
and Stijn Wuyts \altaffilmark{13}
}

\affil{$^1$College of Physical Science and Technology, Shenyang Normal University, Shenyang 110034, China; fengshan@ucolick.org\\
$^2$UCO/Lick Observatory, Department of Astronomy and Astrophysics, University of California, Santa Cruz, CA 95064, USA\\
$^3$Department of Astronomy, University of Massachusetts, Amherst, MA 01003, USA \\
$^4$Aix Marseille Universit$\rm \acute{e}$, CNRS, LAM - Laboratoire d’Astrophysique de Marseille, 13388, Marseille, France \\
$^5$Space Telescope Science Institute, 3700 San Martin Boulevard, Baltimore, MD, 21218, USA \\
$^6$INAF Osservatorio Astronomico di Roma, Via Frascati 33,00040 Monteporzio (RM), Italy \\
$^7$National Astronomical Observatories, Chinese Academy of Sciences, A20 Datun Road, Beijing, 100012, China \\
$^8$Tianjin Astrophysics Center, Tianjin Normal University, Tianjin 300387, China \\
$^9$Observatories of the Carnegie Institution for Science, Pasadena, CA 91101, USA \\
$^{10}$Johns Hopkins University, 3400 N. Charles Street, Baltimore, MD 21218, USA \\
$^{11}$Department of Physics and Astronomy, University of Kentucky, Lexington, KY 40506-0055, USA \\
$^{12}$Department of Physics and Astronomy, Colby College, Mayflower Hill Dr., Waterville, ME 0490, USA\\
$^{13}$Max-Planck-Institut f¨ur extraterrestrische Physik, Postfach 1312, Giessenbachstr., D-85741 Garching, Germany
}

\begin{abstract}
We have made a serendipitous discovery of a massive ($\sim5 {\times}$$10^{11} \rm M_{\odot}$) 
cD galaxy at z=1.096 in a candidate rich cluster in the HUDF area of GOODS-South. 
This brightest cluster galaxy (BCG) is the most distant cD galaxy confirmed to date.
Ultra-deep HST/WFC3 images reveal an extended envelope starting from $\sim 10 \kpc$ 
and reaching $\sim 70 \kpc$ in radius along the semi-major axis. 
The spectral energy distributions indicate that both its inner component 
and outer envelope are composed of an old, passively-evolving (specific star formation 
rate $\rm <10^{-4} Gyr^{-1}$) stellar population. The cD galaxy lies 
on the same mass-size relation as the bulk of quiescent galaxies 
at similar redshifts. The cD galaxy has a higher stellar mass 
surface density ($\rm \sim M_*/R_{50}^2$) but a similar velocity 
dispersion ($\rm \sim \sqrt{M_*/R_{50}}$) to those of more-massive, nearby cDs. 
If the cD galaxy is one of the progenitors of today's more massive cDs, its size ($\rm R_{50}$) 
and stellar mass have had to increase on average by factors of 
$3.4\pm1.1$ and $3.3\pm1.3$ over the past $\sim 8$ Gyrs, respectively. 
Such increases in size and stellar mass without being accompanied 
by significant increases in velocity dispersion are consistent with evolutionary 
scenarios driven by both major and minor dissipationless (dry) mergers. 
If such cD envelopes originate from dry mergers, 
our discovery of even one example proves that some BCGs entered the dry merger phase at epochs earlier than $z = 1$. 
Our data match theoretical models which predict that the continuance of dry mergers at $z<1$ can result in structures 
similar to those of massive cD galaxies seen today. Moreover, our discovery is a surprise given that the extreme depth of the HUDF 
is essential to reveal such an extended cD envelope at $z > 1$ and, yet, the HUDF covers 
only a minuscule region of sky ($\sim 3.1 \times 10^{-8}$). Adding that cDs are rare, our serendipitous discovery hints that 
such cDs may be more common than expected, perhaps even ubiquitous. Images reaching HUDF depths of more area (especially 
with cluster BCGs at $z > 1$) are needed to confirm this conjecture.


\end{abstract}

\section{Introduction}
cD galaxies are the most luminous and most massive galaxies in the local Universe 
and are mostly located close to the centers of galaxy clusters. 
Earlier studies have characterized cD galaxies by their excess light (`envelopes') 
over the de Vaucouleurs (or $r^{1/4}$, de Vaucouleurs 1948) profile 
or S$\acute{\rm e}$rsic (or $r^{1/n}$, S$\acute{\rm e}$rsic 1968) profile 
at large radii \citep[e.g.,][]{mns64,Schombert88,glc+96}. More recent studies have identified cD galaxies using 
improved Petrosian \citep[][]{petrosian76} parameter profiles \citep[][]{brough+05,Patel+06,Liu+08}. 
A large fraction of massive nearby brightest cluster galaxies (BCGs) can be classified as cD galaxies \citep[][]{Liu+08}.

The formation mechanism of BCGs/cDs and their evolution are not well-understood. 
To that end, several formation mechanisms 
including galactic cannibalism \citep[][]{White76,OH77}, tidal 
stripping from cluster galaxies \citep[][]{GO72,Richstoen76,Merritt85} 
and star formation on BCGs by cooling flows \citep[][]{Fabian94} have been proposed. 
Recent semi-analytic models and simulations of galaxy formation suggested 
that BCGs form in a two-phase process: an initial collapse with 
rapid cooling and star formation at high redshift is followed by 
later ($z<1$) growth through multiple dissipationless (dry) 
mergers of pre-existing progenitors (e.g., De Lucia \& Blaizot 2007; 
Ruszkowski \& Springel 2009; Naab et al. 2009; Laporte et al. 2012). 
Very recently, Laporte et al. (2013) showed that BCGs 
can form through dissipationless mergers of quiescent massive 
$z=2$ galaxies in their simulations in a $\rm {\Lambda}CDM$ 
universe, which likely indicates that BCGs entered the dry merger 
phase at epochs earlier than $z=1$. Interestingly, 
they found cD signatures among their simulated BCGs by $z\sim1$ (Laporte et al. 2013). 
Recent observational studies also provide evidence that 
BCGs build up a large part of their stellar masses via dry mergers 
at $z<1$ (e.g., Liu et al. 2008,2009; Tovmassian \& Andernach 2012; 
Lidman et al. 2012). However, other studies argued that BCGs 
evolve minimally since $z\sim1$ (Whiley et al. 2008; Stott et al. 2011) 
and mergers are unlikely to be important for the 
growth of BCGs at $z<1$ (e.g., Ascaso et al. 2011). 
Our early work (Liu et al. 2008) showed that BCGs with dry mergers 
evolve to possess light profiles similar to those of local cD galaxies 
by depositing stellar plumes into the BCG outskirts.

Although the origin of cD galaxies, and the contribution of their envelopes to the intra-cluster light, 
are still not completely clear \citep[e.g.,][]{Gonzalez+05,Zibetti+05}, 
the extended stellar haloes of cD galaxies to surface brightness 
$\mu(r) < 27~\magsec$ in the rest-frame Sloan $r$-band are likely to come from galaxies themselves \citep[][]{Tal11}: 
the intra-cluster light has much lower surface brightness and only 
dominates at large radius \citep[$r \ga$ 80$-$100 $\kpc$; for detailed 
discussions, see][]{Zibetti+05,Bernardi+07,Lauer+07,Tal11}. Even so, it 
is difficult to identify faint cD signatures in the high-redshift 
($z>1$) galaxies due to the cosmological surface brightness dimming. 
Therefore, it is still unknown whether cD galaxies can form at high redshift. 
\citet[][]{Best98} claimed a detection of cD-like signatures 
in two high-redshift 3CR radio galaxies (3C 22 at z=0.936 and 3C 41 at z=0.795) based on shallow (one orbit) HST/WFPC2 imaging data. 
However, their results were not confirmed by deeper imaging observations. 

In this paper, we report the serendipitous discovery of a massive cD galaxy (GOODS-S J033237.19-274608.1, named after GOODS-S and 
its J2000 coordinate: RA=53.1549542, DEC=-27.7689028) at z=1.096 (see \S\ref{sec:epsilon_s}) in a candidate rich cluster in the HUDF area of GOODS-South. 
We use its ultra-deep HST/WFC3 mosaic image to $\sim 30 \magsec$ to confirm the existence of an extended cD envelope 
characterized by distinct Petrosian parameter profiles. 
We also investigate the stellar population and environment of this distant cD galaxy, 
and explore its possible evolution to today by comparing its structural parameters to those of more-massive, nearby cD galaxies. 
Throughout the paper we adopt a cosmology with a matter density 
parameter $\Omega_{\rm m}=0.3$, a cosmological constant 
$\Omega_{\rm \Lambda}=0.7$ and a Hubble constant of ${\rm H}_{\rm 
0}=70\,{\rm km \, s^{-1} Mpc^{-1}}$. All magnitudes are in the AB 
system unless otherwise stated.

\section{Image Data}

\begin{figure*}
        \centering
        \includegraphics[angle=0,width=0.6\textwidth]{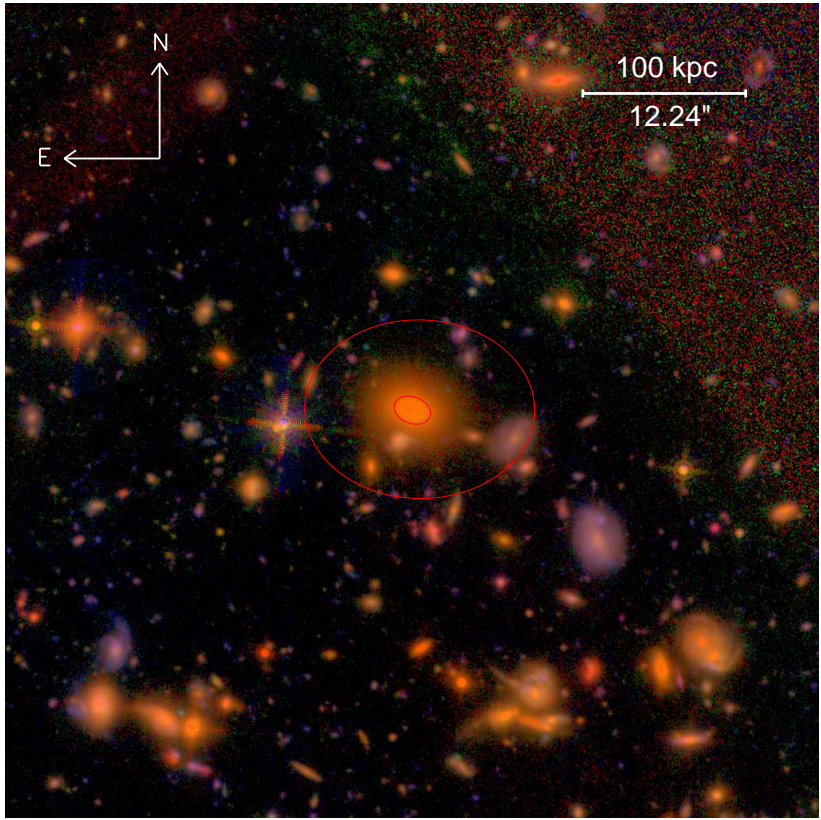}
       \caption{
The pseudo-color image ($\rm 0.5~Mpc \times 0.5~Mpc$) of GOODS-S J033237.19-274608.1. North is up and east is to the left. 
The transitional isophote (inner) and the maximum measured isophote (outer) are marked with red ellipses.
\label{fig:image}}
\end{figure*}

\begin{figure*}
        \centering
        \includegraphics[angle=0,width=0.85\textwidth]{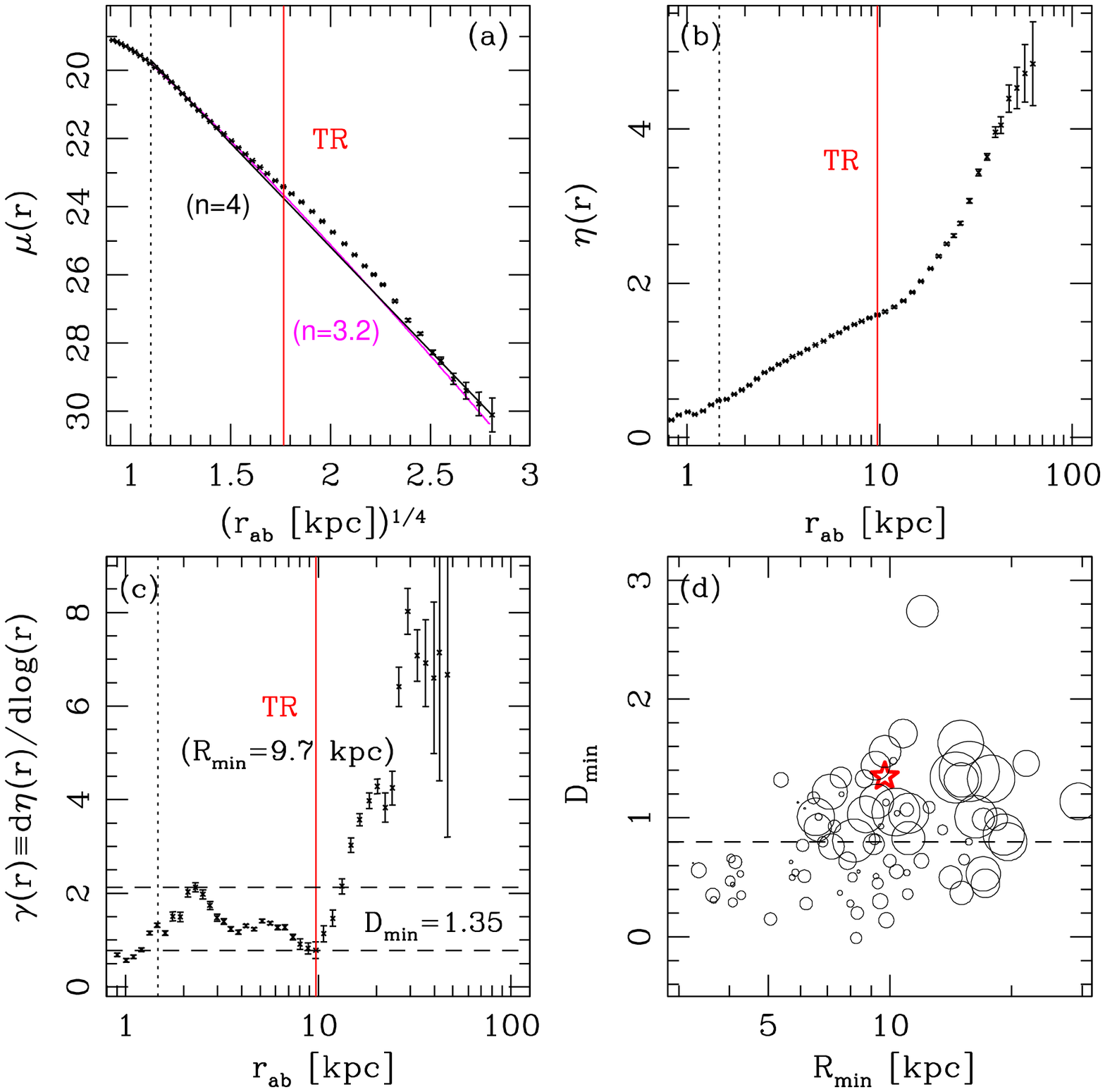}
       \caption{
The panels (a)-(c) show the surface brightness profile (a), the corresponding Petrosian $\eta(r)$ profile (b) 
and $\gamma(r)$ profile (c) in the $J$-band, respectively. The vertical dotted lines mark twice the PSF radius. 
The vertical solid lines mark the transitional radius (TR). 
The solid curves in panel (a) are the best-fit de Vaucouleurs (black) and S$\acute{\rm e}$rsic (magenta) profiles fit to the data 
between twice the PSF radius and maximum measured radius.
Panel (d) illustrates the position of GOODS-S J033237.19-274608.1 on the relation of $\Dmin$ versus $\rmin$ for nearby BCGs in Liu et al. (2008). 
Nearby BCGs are shown with circles and GOODS-S J033237.19-274608.1 is marked with a red star. 
Symbol sizes are proportional to the stellar masses that are derived from the SED fitting (see \S\ref{sec:evolu}).
The horizontal dashed line shows the median value (0.8) of $D_{min}$ in nearby BCGs
\label{fig:sbp}}
\end{figure*}

The $z\sim1.1$ cD galaxy, GOODS-S J033237.19-274608.1, is located 
in the Hubble Ultra Deep Field 
\citep[HUDF,][]{Beckwith+06} region of the south field of the Great 
Observatories Origin Deep Survey \citep[GOODS-S,][]{Giavalisco+04}. 
The HUDF is covered by the HST/WFC3 F105W($Y$), F125W($J$), and F160W($H$) bands imaging from 
G. Illingworth's HUDF09 program \citep[GO 11563,][]{Bouwens+10} as 
well as the Cosmic Assembly Near-IR Deep Extragalactic Legacy 
Survey \citep[CANDELS,][]{Grogin+11,Koekemoer+11}. The HUDF has 
also been observed by the HST/ACS F435W($B$), F606W($V$), F775W($i$), 
and F850LP($z$) bands imaging from \citet{Beckwith+06} 
and GOODS \citep{Giavalisco+04}. We made use of all HST observations 
published prior to June 2012 in the HUDF to maximize our ability 
to detect the low surface brightness envelope of this cD galaxy at 
large radius. These HST mosaics reach the 5$\sigma$ limiting depths 
(0\farcs35-diameter aperture) of 29.7, 30.1, 29.9, 29.4, 29.6, 29.9, 
and 29.9 AB in $B$, $V$, $i$, $z$, $Y$, $J$, and $H$-bands. 
Low-resolution UV (U-band from both CTIO/MOSAIC and VLT/VIMOS) and 
IR (VLT/ISAAC Ks and Spitzer/IRAC 3.6, 4.5, 5.8, 8.0 ${\mu}m$) 
images are also available for this galaxy. Uniform multi-wavelength photometry 
from the UV to 8.0 ${\mu}m$ with the resolution and blending of each band 
taken into account is provided by Guo et al. (2013).

\section{Identification of cD signature}

To quantitatively determine how a surface brightness profile deviates 
from a de Vaucouleurs or S$\acute{\rm e}$rsic profile, we use an objective method based 
on the Petrosian $\eta(r)$ profiles \citep[][]{Liu+08}, defined as
\begin{equation}
\eta(r) \equiv \mu(r) - \langle \mu(r) \rangle,
\label{eq:eta}
\end{equation}
where $\mu(r)$ is the surface brightness in magnitudes at radius 
$r$ and $\langle \mu(r) \rangle $ is the mean surface brightness 
within $r$ \citep[][]{petrosian76}. We also calculate the gradient 
of $\eta(r)$, $\gamma(r) \equiv d\eta(r)/d\log(r)$. 
Liu et al. (2008) have shown that there is always a distinct signature of a 
plateau in the Petrosian $\eta(r)$ profiles and a valley in the $\gamma(r)$ profiles for cD-like BCGs 
with an extended stellar halo (see their Fig.~5). Such signatures 
are not present in normal elliptical galaxies, which are usually 
well-fit by a $r^{1/n}$ profile. We label the radius of the 
minimum in the $\gamma(r)$ profile as $\rmin$. We also define a 
depth, $\Dmin$, which is the difference between the minimum and 
maximum of $\gamma(r)$ outside twice the seeing radius and 
$\rmin$. $\rmin$ is, by definition, the transitional radius (TR) 
between the inner component and outer halo in a cD-like galaxy. 
Our method indicates that the surface brightness deviations of 
overall BCGs from the single S$\acute{\rm e}$rsic profile does not 
have any sharp transitions. However, the more significant the 
deviations of the surface brightness profiles from a single S$\acute{\rm e}$rsic profile (i.e., the more prominent the stellar 
halo in the outskirts), the more distinct the signatures 
(plateau and valley) become statistically (see Liu et al. 2008 for details).

Fig.~\ref{fig:image} is the pseudo-color composite image of GOODS-S J033237.19-274608.1 
from the ACS $V$-band, WFC3 $J$- and $H$-band images. 
The transitional isophote and the maximum measured isophote are marked with red ellipses. 
Panels (a)-(c) of the Fig.~\ref{fig:sbp} show the observed surface brightness profile, the 
corresponding Petrosian $\eta(r)$ profile and $\gamma(r)$ profile 
in the $J$-band, all of which are measured from the mosaic image of CANDELS/Deep + HUDF09 (5 orbits + 34 orbits). 
All radial profiles use the equivalent radius of an ellipse, 
$r_{ab}{\equiv}\sqrt{ab}$, where $a$ and $b$ are the lengths of 
semi-major and semi-minor axes of the ellipse. Contaminating objects 
detected with counts above 1 sigma noise have been masked during our measurement following Liu et al. (2008). 
The F125W ($J$) band is used because it is closest to the rest-frame Sloan $r$-band at this redshift. 
It is clear that the surface brightness profile of GOODS-S J033237.19-274608.1 deviates significantly from 
a single de Vaucouleurs ($r^{1/4}$) profile or S$\acute{\rm e}$rsic ($r^{1/n}$, 
$n\sim3.2$ here) profile, and the $\eta(r)$ profile shows a plateau and 
the $\gamma(r)$ profile shows a valley around the transitional radius. The transitional radius 
$\rmin$ is $9.7 \kpc$ and the corresponding $\Dmin$ value is 1.35.

We investigate the position of GOODS-S J033237.19-274608.1 on a plot of 
$\Dmin$ versus $\rmin$ compared to a representative sample of nearby BCGs in Liu et al. (2008), 
as shown in the panel (d) of Fig.~\ref{fig:sbp}. 
Liu et al. (2008) statistically showed that the larger $\Dmin$ and the transitional radius 
$\rmin$ of a BCG are, the more significantly extended its envelope appears to be. 
BCGs with $\Dmin$ larger than 0.8 are more likely to be classified as cD galaxies. 
Note that $\Dmin=0.8$ is the median value for all nearby BCGs. Our galaxy, 
GOODS-S J033237.19-274608.1 at $z=1.096$, has $\Dmin=1.35$ and is therefore termed a cD galaxy. 
The stellar envelope of this high-redshift cD galaxy starts to 
dominate from $r_{ab}=9.7 \kpc$ (the corresponding semi-major axis 
radius is $a=11.6 \kpc$) and extends to $65.3 \kpc$ ($a=70.12 \kpc$).

\section{The size and stellar mass}

It is unreasonable to characterize the sizes of cD galaxies by half-light radii (effective radii) 
from single S$\acute{\rm e}$rsic ($n=4$ for de Vaucouleurs) fits since the surface brightness profiles of cD galaxies 
usually show strong deviations from single S$\acute{\rm e}$rsic profiles. We opt to use the real 
isophotal half-light radius to characterize the size of this cD galaxy following Liu et al. (2008). 
The derived half-light radius of GOODS-S J033237.19-274608.1 within the maximum 
measured isophote in the $J$-band is only $\sim5.25\kpc$. In order 
to estimate the effects of point-spread-function  (PSF) on the size measurement, 
we first used the GALFIT package \citep[][]{Peng+02} 
to fit a conventional two-dimensional model, convolved with the 
real PSF, to the observed image. A double-S$\acute{\rm e}$rsic 
model was applied to determine the best-fit model image. We then 
added the residual image to the best-fit model image (PSF 
deconvolved) to correct the uncertainty of model fitting. It should be noted that 
the residual image is not deconvolved for PSF. However, this 
does not significantly affect the accuracy of our method since the 
flux ratio of residual-to-observed images is only $\sim3\%$. We did 
the isophotal photometry and half-light radius measurement for the 
`residual-corrected' model image in the same way as for the 
observed image. The obtained half-light radius is $\sim5.19\kpc$, which shows the PSF effect is not significant.

We followed Guo et al. (2012) to fit spectral energy distributions (SEDs) composed of 
homogeneous aperture magnitudes from the UV to $8{\mu}m$ to estimate the stellar mass. 
We used an updated version (CB07) of \citet[][]{bc03} stellar population synthesis models with a range of stellar population properties 
and dust extinctions. We assumed  models with a solar metallicity, a Salpeter \citep[][]{Salpeter55} initial mass function (IMF), and a \citet[][]{Calzetti+00} law. 
The derived stellar mass is $\rm  10^{11.62\pm0.08}M_{\odot}$. It becomes $\sim5 {\times}$$10^{11} \rm M_{\odot}$ ($10^{11.71\pm0.08} \rm M_{\odot}$) 
after accounting for the small difference between the used aperture magnitude and isophotal magnitude within the maximum measured ellipse in the $J$-band.

\section{Optical spectroscopy and stellar population}\label{sec:epsilon_s}

\begin{figure}
        \centering
        \includegraphics[angle=0,width=0.49\textwidth]{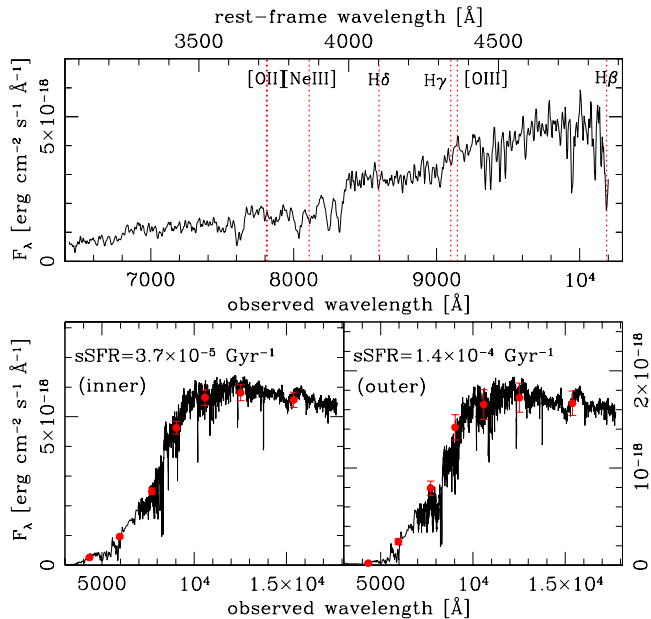}
        \caption{
The top panel is the optical spectrum of GOODS-S J033237.19-274608.1 observed by the VLT/FORS2 instrument.
The bottom panels are the best-fit SEDs composed of high-resolution HST bands for the inner component
(bottom left) and the outer envelope (bottom right), respectively.
\label{spec_sed}}
\end{figure}

We show the optical spectroscopy of GOODS-S J033237.19-274608.1 in the top 
panel of Fig.~\ref{spec_sed}, which is observed by the VLT/FORS2 
instrument in the ESO/GOODS spectroscopic program. The grating used 
has a scale of roughly 3.2 $\rm \AA/pixel$ and a nominal resolution of $\lambda/\Delta{\lambda}=860$. 
The whole spectral range is $\rm 6000-10800\AA$ \citep[see][for details]{Vanzella+05}. 
The derived spectroscopic redshift of GOODS-S J033237.19-274608.1 is $1.096\pm0.001$. 
The spectroscopy of this galaxy shows weak emission lines (e.g., [OII]3727) 
and a weak $\rm H{\delta}$ absorption line, indicating that it is dominated by an old stellar population.

We fitted the SEDs of the inner component ($r<R_{min}$) and outer envelope ($R_{min}<r<30\kpc$) to investigate each of their stellar 
populations. The unmasked imaging data in seven 
high-resolution HST bands ($B$,$V$,$i$,$z$,$Y$,$J$,$H$) were used. 
The cut off at $30\kpc$ for the outer region is to ensure 
high signal-to-noise ratios for all bands at large radius. The 
results are shown in the bottom panels of Fig.~\ref{spec_sed}. 
The derived specific star formation rates (sSFRs) for the inner 
component and outer envelope are $\rm 3.7 \times 10^{-5} Gyr^{-1}$ 
and $\rm 1.4 \times 10^{-4} Gyr^{-1}$, respectively. 
The sSFRs, i.e. the SFR per unit stellar mass, were calculated by dividing the SFRs by the stellar masses, 
which were derived from the SED fittings separately. The sSFRs of both components are consistent with 
being passively-evolving stellar populations.

\section{The environment}

\begin{figure}
\begin{center}
\begin{minipage}[t]{0.92\linewidth}
\centerline{\includegraphics[angle=0,width=1.0\textwidth]{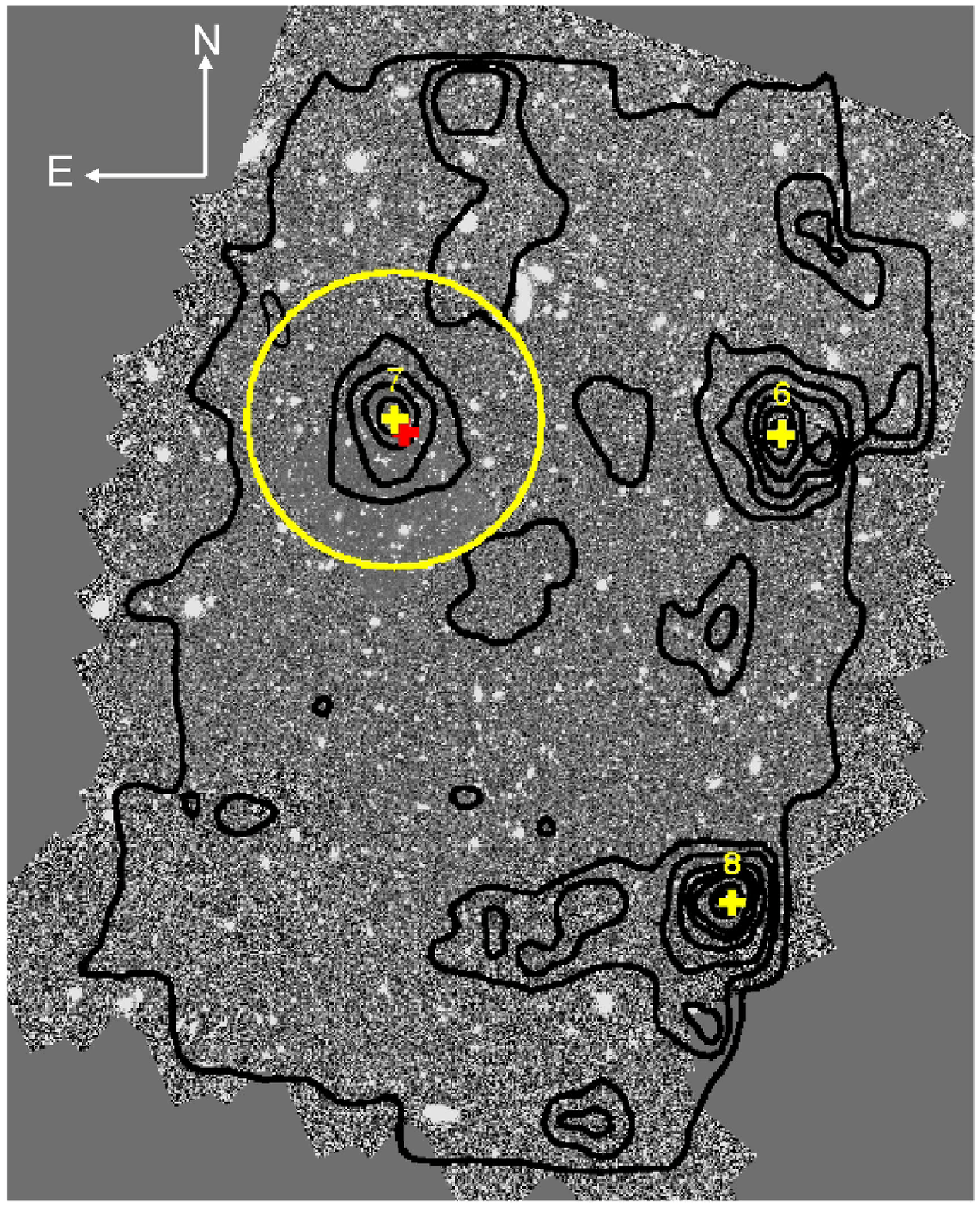}}
\end{minipage}%
\begin{minipage}[t]{0.92\linewidth}
\centerline{\includegraphics[angle=0,width=1.0\textwidth]{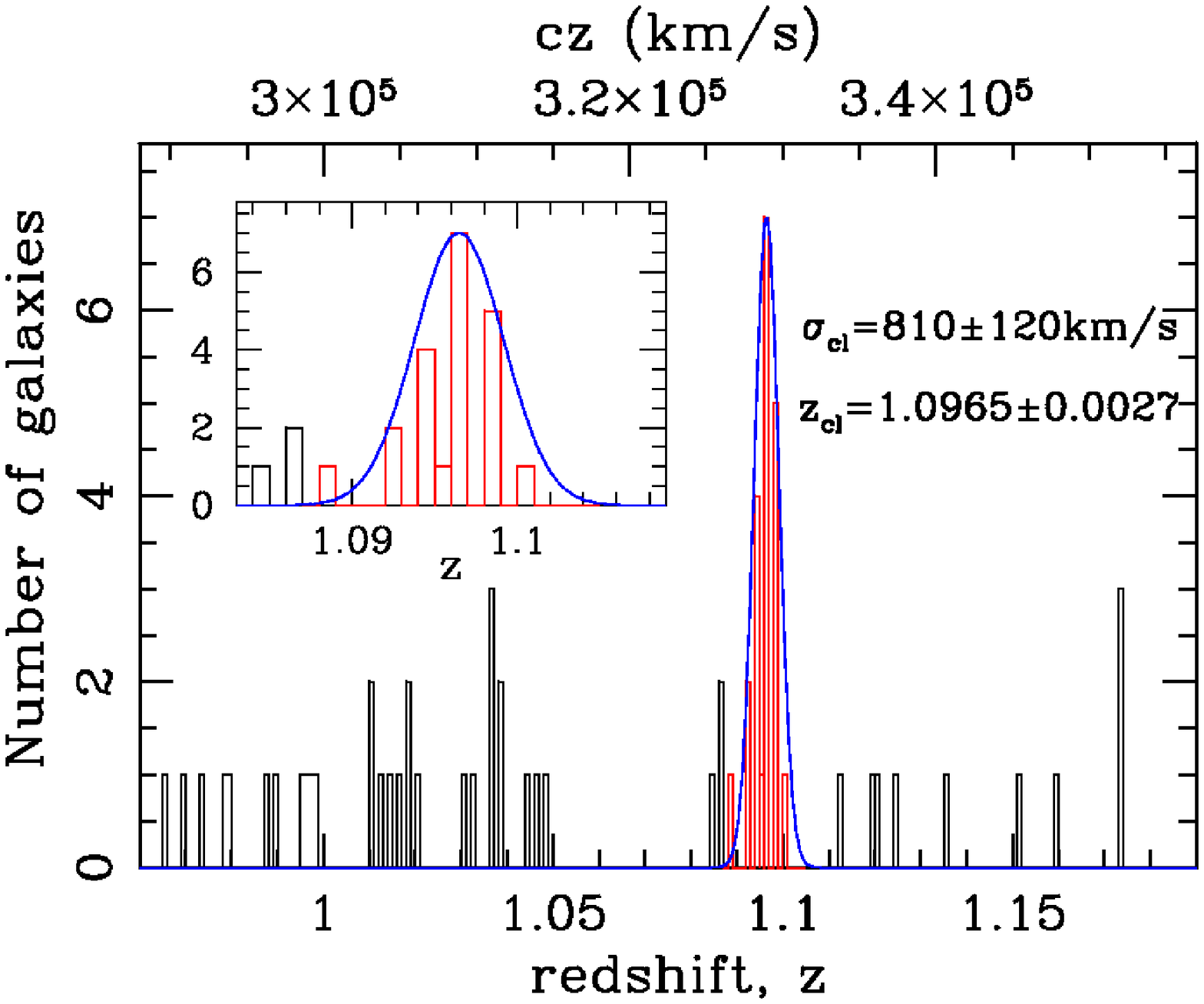}}
\end{minipage}%
\end{center}
\caption{
The top image illustrates the position of GOODS-S J033237.19-274608.1 (red cross) on the WFC3 $J$-band image of the entire GOODS-S field. 
The projected density isosurfaces for structures at $z_p\sim1.05$ identified by \citet[][]{Salimbeni+09} are superimposed on it. 
Yellow crosses indicate the density peak of each structure, and the number of each is the ID of the structure given by \citet[][]{Salimbeni+09}. 
The big yellow circle marks the projected range within 1 Mpc from the density peak of the structure with ID=7.
The bottom panel shows the distribution of redshifts (radial velocities) of galaxies with secure spectroscopic redshifts 
near the cluster redshift and within the marked range (yellow circle). The curve is the final best-fit Gaussian distribution. 
Red histograms represent the data used during the final fitting. Black histograms represent galaxies which 
are likely outliers rejected by our iterative fittings. The inset is designed to show the distribution of spectroscopically confirmed members more clearly.
} \label{cluster_iden}
\end{figure}

It is well known that cD galaxies in the local Universe are usually BCGs and are mostly found in the centers of rich clusters. 
We now investigate whether this high-redshift cD galaxy is also located in a cluster environment. 
\citet[][]{Salimbeni+09} have found three group/cluster structures around the photometric redshift $z_p\sim1.05$ in the entire GOODS-S field. 
The projected density isosurfaces of these three structures are shown in the top image of Fig.~\ref{cluster_iden} (also see 
panel c in Salimbeni et al. 2009). It is obvious that the discovered $z\sim1.1$ cD galaxy is very close to the projected density peak of the structure with ID=7. 
\citet[][]{Salimbeni+09} estimated that this structure has a photometric redshift $z_p\sim1.04$, $\rm r_{200} \sim 1.2 - 1.8 Mpc$ and 60 member galaxies. 
Our cD galaxy is the most massive galaxy within the redshift gap of $\Delta z_p=1.04\pm0.1$ in this structure.
We searched for other galaxies with secure spectroscopic redshifts (quality flag = 1 or 2) 
in the literature \citep[][]{Vanzella+05,Vanzella+06,Balestra+10,Cooper+12} 
within a projected distance of 1 Mpc from the density peak of this structure. 
The distribution of redshifts (radial velocities) of these galaxies 
(with a bin size of 0.001, corresponding to typical errors of spectroscopic redshifts) 
near the estimated structure redshift is shown in the bottom panel of Fig.~\ref{cluster_iden}. 
We iteratively fitted this distribution with a Gaussian function and rejected all outliers which lie 
beyond 3 sigma of the best-fit Gaussian distribution. 
The final best fit includes 21 objects all within 3 sigma of the best-fit Gaussian, 
with an inferred velocity dispersion of 810$\pm$120km/s. 
We derived a precise redshift location of $z=1.0965\pm0.0027$ for this structure. 
These values are consistent with the results (797$\pm$138km/s and $z=1.0968\pm0.0027$) estimated using the biweight estimators, 
computed with the Rostat package (Beers et al. 1990). 
The redshift of the cD galaxy places at the center of the cluster redshift distribution. 
These properties collectively suggest that GOODS-S J033237.19-274608.1 
is both likely to be in a cluster environment and is also located close to the cluster center. 
Future spectroscopic observations can confirm more cluster galaxies and quantify the richness of this potential rich cluster.

\section{Summary and Discussion}  \label{sec:evolu}

\begin{figure*}
\begin{center}
\begin{minipage}[t]{0.88\linewidth}
\centerline{\includegraphics[angle=0,width=1.0\textwidth]{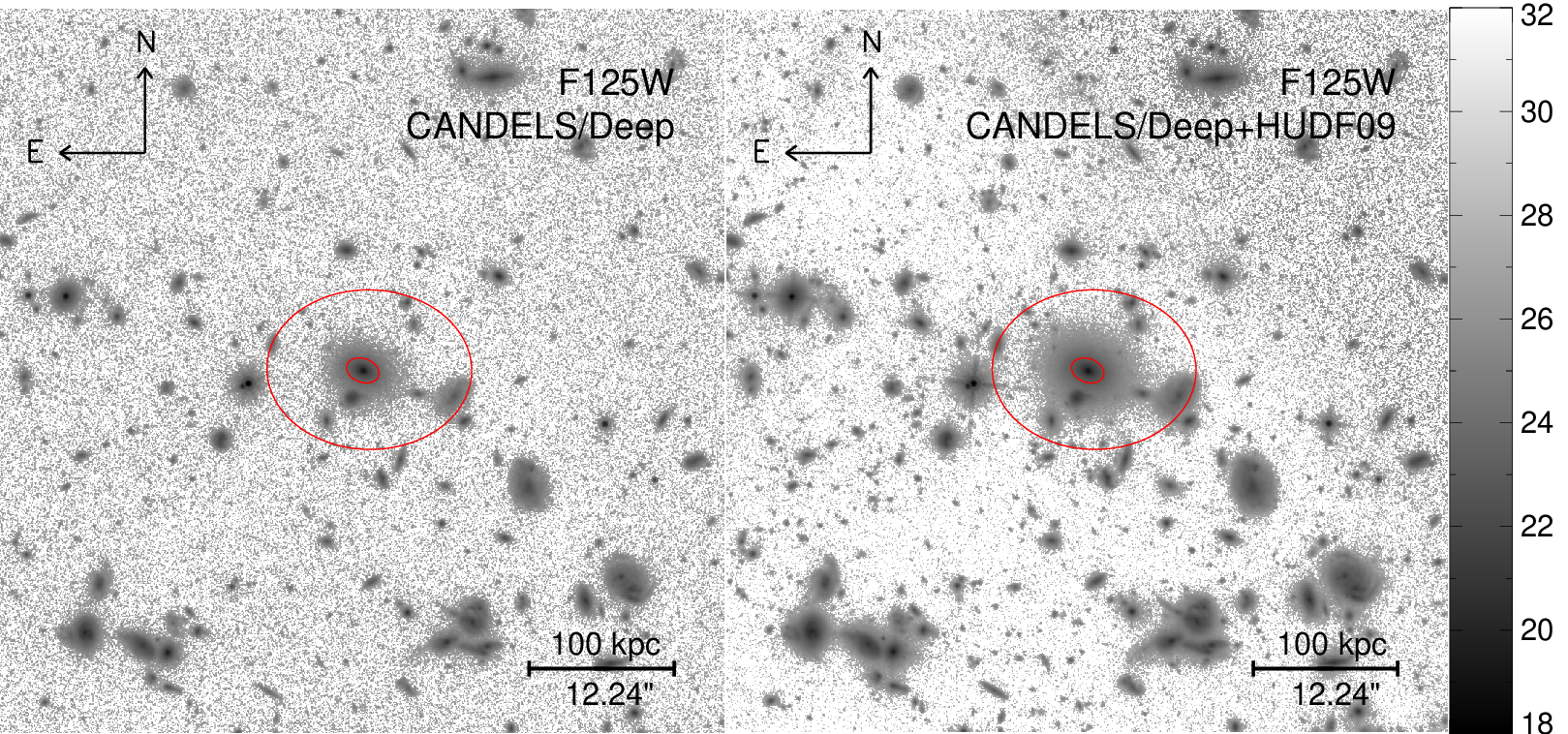}}
\end{minipage}%
%\hspace{8mm}
%\\
\begin{minipage}[t]{0.5\linewidth}
\centerline{\includegraphics[angle=0,width=1.0\textwidth]{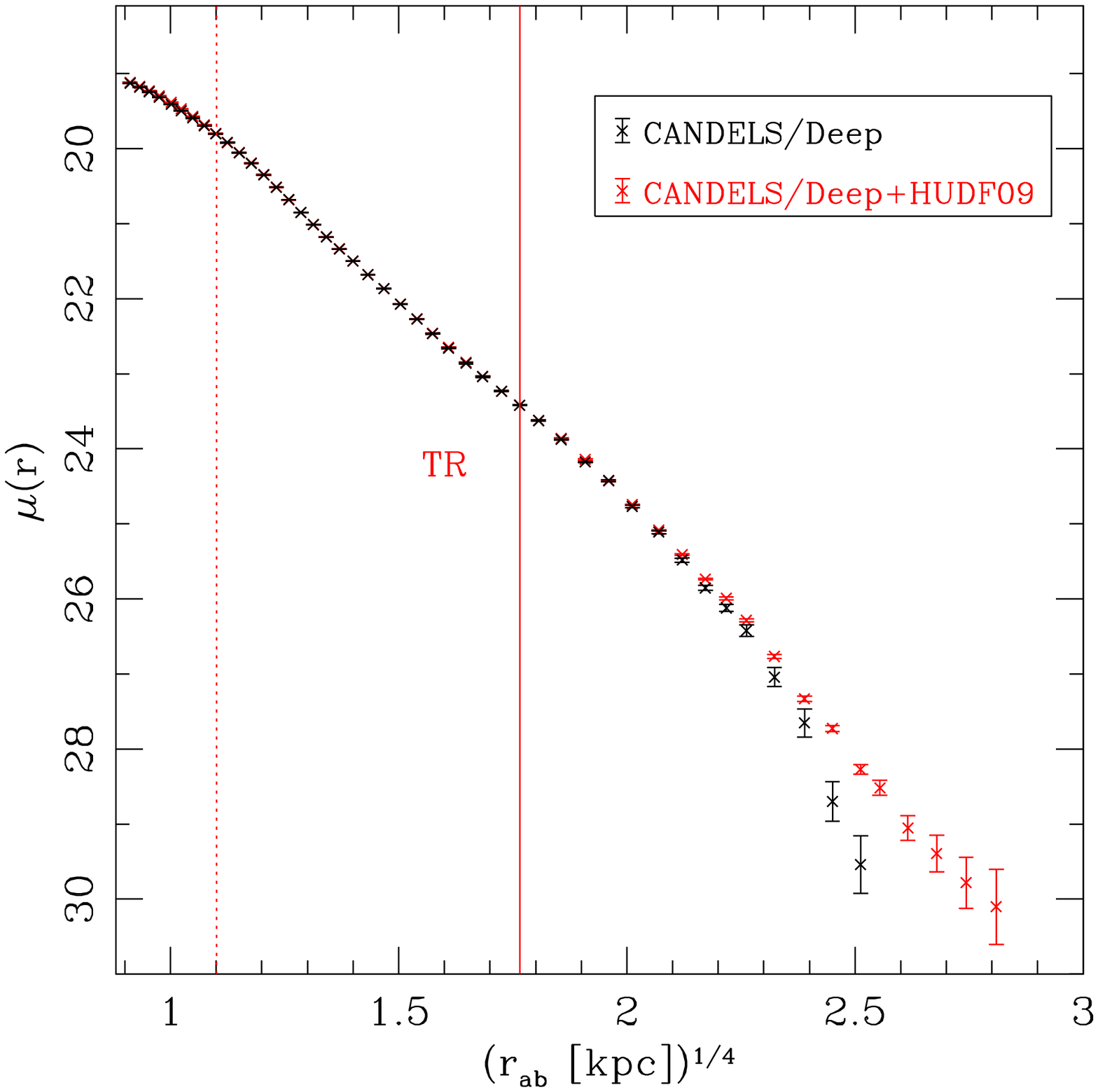}}
\end{minipage}%
\end{center}
\caption{ Comparisons of the $J$-band images (top) and surface brightness profiles (bottom) 
between the CANDELS/Deep only and the CANDELS/Deep + HUDF09 observations. The sizes of images are the same as that in Fig.~\ref{fig:image}. 
North is up and east is to the left. 
The transitional isophote (inner) and the maximum measured isophote (outer) are marked on each image with red ellipses. 
%The transitional isophote is marked on each image with a red ellipse. 
The transitional radius (TR) is also marked on the bottom profiles with a vertical solid line. 
The vertical dotted line in the bottom panel marks twice the PSF radius.
} \label{candels_hudf09}
\end{figure*}

\begin{figure*}
        \centering
        \includegraphics[angle=0,width=0.85\textwidth]{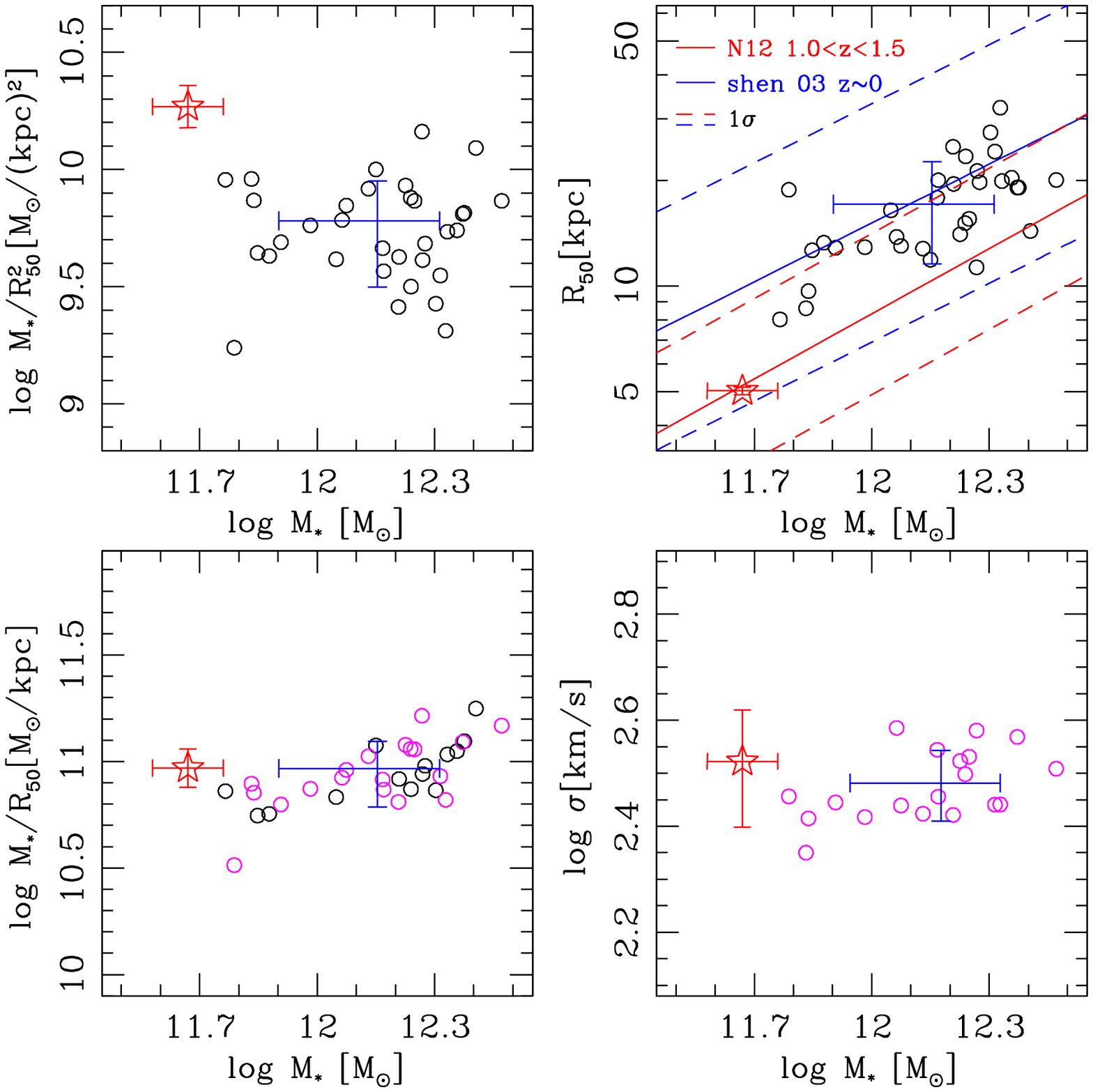}
       \caption{
The stellar mass surface densities ($\rm \sim M_*/R_{50}^2$), sizes ($\rm R_{50}$), 
$\rm M_*/R_{50}$ and velocity dispersions ($\sigma$) as a function of stellar masses for the high-redshift cD galaxy (star) 
and more-massive, nearby cDs (circles). Our mass-size relation in the top right panel is overlaid with 
the mass-size relations of Newman et al. (2012) in the redshift range $1.0<z<1.5$ and Shen et al. (2003) at $z\sim0$. 
19 out of 33 nearby cDs have measured velocity dispersions values. These objects are shown with magenta circles 
in the bottom panels. The blue error bars in each panel show the mean values and the corresponding standard deviations of parameters of nearby cDs. 
The errors of data points are shown only for the high-redshift cD galaxy for clarity.
\label{fig:size-smass}}
\end{figure*}

%%%%%%%%%%%%
We have made a serendipitous discovery that the most massive ($\sim5 {\times}$$10^{11} \rm M_{\odot}$) 
galaxy (GOODS-S J033237.19-274608.1), i.e., the BCG,  in a candidate rich cluster at z=1.096  in the HUDF area of GOODS-S is a cD,
with an outer envelope extending in radius to $\sim 70 \kpc$ along the semi-major axis. 
This is the most distant, massive cD galaxy confirmed to date.
The HUDF image is unique in being the deepest and, thus, as seen in 
Fig.~\ref{candels_hudf09}, such a faint, extended, stellar envelope 
would not have been otherwise visible. 
Moreover, since the HUDF only covers less than 5 square arcminutes (i.e., 1 part in 30 millionth of the sky)
and BCGs are expected to be very rare at $z>1$, our discovery of even one such BCG being a cD 
was a total surprise and is certainly fortuitous by any expectation. 
Beyond the following discussion of the implications for the early and later formation of some cDs, 
our discovery hints that many other $z>1$ BCGs might be found 
with such extended cD halos, if more areas or clusters were covered with deep enough images.

A clue to the early formation of cDs comes from the SEDs of both the inner component and outer envelope of this cD galaxy; 
the SEDs match those of old, passively-evolving stellar populations. 
Assuming that such cD envelopes originate from dry mergers (see Liu et al. 2008), 
our discovery of even one such example at $z > 1$ proves that some BCGs entered the dry merger phase at epochs earlier than $z=1$, 
a result consistent with recent simulations by Laporte et al. (2013).

Next, we explore the possible evolution of our $z > 1$ cD galaxy to today
by comparing its structural parameters (i.e., mass, size, and velocity dispersion) to those of the more massive cDs in the local Universe. 
The stellar masses of local cDs were derived using 
the same CB07 models with the same Salpeter IMF, solar metallicity, and a Calzetti extinction law (Calzetti et al. 2000) 
as adopted  for deriving the mass of our $z > 1$ cD galaxy. The SEDs used Petrosian magnitudes in $ugriz$ bands from the Sloan Digital Sky Survey and the extrapolated 
magnitudes in $JHK_s$ bands from the Two Micron All Sky Survey 
Extended Source Catalog (2MASS XSC).  
The final masses were re-scaled by adding the differences between the Sloan 
$r$-band Petrosian magnitudes and isophotal magnitudes measured within $25\magsec$.

The $r$-band isophotal half-light 
radius of nearby BCGs within $25\magsec$ were measured 
by Liu et al. (2008). To make the comparison at the same rest-frame bandpass, we 
applied the best-fit global SEDs of GOODS-S J033237.19-274608.1 to convert the 
observed $J$-band surface brightness profile to the rest-frame 
Sloan $r$-band. The surface brightness profile has also been corrected 
for the cosmological dimming of 10 log (1+z). The corresponding stellar mass and half-light 
radius of GOODS-S J033237.19-274608.1 within $25\magsec$ in the rest-frame $r$-band are 
$10^{11.67\pm0.09}\rm M_{\odot}$ and $5.02\pm 0.12\kpc$, respectively. 
These errors include both random errors and errors from the band shifting and PSF effects. 
There are 64 nearby BCGs with log $\rm M_{*}\geq11.67$ out of the 85 BCGs in Liu 
et al. (2008), 33 of which have $\Dmin>0.8$ and can be classified as cD galaxies with high probability.
19 of those 33 nearby cDs have the measured velocity dispersions available in Liu et al. (2008). 
We derived a velocity dispersion of $\rm 333\pm83km/s$ for GOODS-S J033237.19-274608.1 from its optical spectroscopy, 
which is consistent with the value ($\rm 324\pm32km/s$) estimated by \citet[][corresponding 
to ID CDFS-18 in their paper]{vanderWel+05}. We show the stellar mass surface 
densities (${\sim}\rm M_*/R_{50}^2$)\footnote{The stellar mass surface density is actually is $\rm M_*/2{\pi}R_{50}^2$, but we omit the constants.}, 
sizes ($\rm R_{50}$), $\rm M_*/R_{50}$\citep[sometimes used to infer velocity dispersion, see][]{Franx+08}
\footnote{The true stellar velocity dispersion is $\rm \sigma^2 \propto GM/R_{50}$, where M is the total
mass including stars, gas, and dark matter. Franx et al. (2008) provide a value
of the coefficient through the fitting of a sample of SDSS galaxies:
$\rm \sigma^2 =0.3GM_*/R_{50}$.
}, and velocity dispersion ($\sigma$) as a function of stellar masses of this high-redshift cD galaxy and 
more-massive, nearby cDs in Fig.~\ref{fig:size-smass}, respectively.

The distant cD galaxy clearly has a higher stellar mass surface density (i.e., more compact) but 
a similar velocity dispersion ($\rm \sim \sqrt{M_*/R_{50}}$) to those of more-massive, nearby cDs.
It lies, however, on the same mass-size relation as the bulk of quiescent galaxies at similar redshift ($1.0<z<1.5$), as derived by Newman et al. (2012). 
The more-massive, nearby cDs, in contrast, lie on the $z\sim0$ mass-size relation (within the scatter) derived by Shen et al. (2003) who used Petrosian-based quantities. 
To evolve the distant cD galaxy into one of the more massive nearby cDs, its size and 
stellar mass have to increase, on average, by factors of $3.4\pm1.1$ and $3.3\pm1.3$ over the past $\sim 8$ Gyrs. 

As mentioned in the introduction, recent studies from both 
numerical simulations and observations show that BCGs accumulate a 
large part of their stellar mass via dry mergers at $z<1$. 
Ruszkowski \& Springel (2009) first investigated the effect of 
dry mergers on the scaling relations of BCGs in simulations. 
Laporte et al. (2012) argued, however, that the stellar masses and 
sizes of the final merger remnants derived from the Ruszkowski \& Springel simulations 
were too massive ($\rm \sim10^{13} M_{\odot}$) and large ($\rm \sim100 \kpc$) 
compared to real BCGs at $z\sim0$. They thus revised the simulations 
of Ruszkowski \& Springel by including the effects of initial compactness 
on the subsequent evolution of the mass-size relation of the BCGs (Laporte et al. 2012). 
The revised simulations showed that the size evolution of the most extended BCGs (e.g., cDs) 
at high redshift scale linearly with stellar mass ($R_{50}{\propto} M_{*}$).
\citet[][]{Nipoti+09} also investigated the dry-merging evolution scenario based on a set of simulations 
and showed that major and minor dry mergers increase half-light radius with 
stellar mass as $R_{50}{\propto} M_{*}^{1.09\pm0.29}$ but have 
little effects on velocity dispersion ($\sigma {\propto} M_{*}^{0.07\pm0.11}$). 
The reasonable agreement between our observations 
and theoretical predications mentioned above shows that the continuance of dry mergers 
at $z<1$ can result in structures similar to those of massive cD galaxies seen today and thus
support the view that dry mergers play an important role in the late evolution of cD galaxies.

%%%%%%%%%%%%%%

\acknowledgments
We thank Steve Willner, Michael Cooper, and Henry Ferguson for useful comments, especially thank Sara Salimbeni for her data on the large-scale structure. 
We also acknowledge the anonymous referee for an expert and valuable report that improved the paper.
This project was supported by an NSF grant, AST 08-08133, and the National Science Foundation of China (11103013,10973011). 
F. S. Liu thanks S. M. Faber and D. Koo for hosting his visit at UCSC.

\end{document}